\documentclass{article}

\usepackage{times}
\usepackage[pdftex]{graphicx}
\usepackage{amsmath}
\usepackage{amssymb}
\usepackage{authblk}
\usepackage{bm}
\usepackage{tabularx}
\usepackage{multirow}
\usepackage{algorithm}
\usepackage{algorithmic}

\usepackage{fmt}
\usepackage{mcr}
\usepackage{pkg}
\usepackage{thmE}
\usepackage{algE}

\usepackage[pagebackref=true,breaklinks=true,letterpaper=true,bookmarks=false]{hyperref}
\usepackage[margin=1in]{geometry}

\begin{document}

\title{Transformer-based Personalized Attention Mechanism for Medical Images with Clinical Records}

\author[1]{Yusuke Takagi}
\affil[1]{Department of Computer Science, Nagoya Institute of Technology}
\author[2]{Noriaki Hashimoto}
\affil[2]{RIKEN Center for Advanced Intelligence Project}
\author[1]{Hiroki Masuda}
\author[3]{Hiroaki Miyoshi}
\affil[3]{Department of Pathology, Kurume University School of Medicine}
\author[3]{Koichi Ohshima}
\author[1]{Hidekata Hontani}
\author[2,4,*]{Ichiro Takeuchi}
\affil[4]{Department of Mechanical Systems Engineering, Nagoya University\protect\\{\it  ichiro.takeuchi@mae.nagoya-u.ac.jp}}

\date{}
\maketitle

\begin{abstract}
  In medical image diagnosis, identifying the attention region, i.e., the region of interest for which the diagnosis is made, is an important task.
  Various methods have been developed to automatically identify target regions from given medical images.
  However, in actual medical practice, the diagnosis is made based not only on the images but also on a variety of clinical records.
  This means that pathologists examine medical images with some prior knowledge of the patients and that the attention regions may change depending on the clinical records.
  In this study, we propose a method called the \emph{Personalized Attention Mechanism (PersAM)}, by which the attention regions in medical images are adaptively changed according to the clinical records.
  The primary idea of the PersAM method is to encode the relationships between the medical images and clinical records using a variant of Transformer architecture.
  To demonstrate the effectiveness of the PersAM method, we applied it to a large-scale digital pathology problem of identifying the subtypes of 842 malignant lymphoma patients based on their gigapixel whole slide images and clinical records.

\end{abstract}

\section{Introduction}
Medical images are often diagnosed on the basis of specific regions of interest in the images rather than their entirety.
For example, cancer pathologists typically focus on specific tumor regions rather than the entire pathological tissue specimen.
In this study, we refer to such regions as \emph{attention regions}.
Developing computational methods to estimate the attention regions is an important task in medical image analysis to obtain high performance and explainability.
In existing methods, the attention regions are predominantly estimated based solely on the images themselves~\cite{ilse2018attention,hashimoto2020multi,li2021dual}.
However, in clinical practice, pathologists use both the imaging information and various clinical records (including basic demographic details such as patient age and gender, the results of various medical examinations, and genetic information).
It is well-recognized among pathologists that patient-specific information can help them focus on specific tissues in the specimens or narrow the diagnostic target classes.
In practice, the region to be focused on in a tissue slide changes depending on the type of organs from which a tissue specimen is sliced, or the results of a medical interview and some medical tests narrow down suspected diseases.
It is known that the additional use of clinical record information can enhance the performance also in medical image analysis~\cite{yala2019deep,yap2018multimodal,thung2017multi,nie2019multi}.
In this study, we introduce a framework, called the \emph{Personalized Attention Mechanism (PersAM)} framework, that adaptively changes the attention regions in medical images according to patient-specific information.
The PersAM framework mimics pathologists' decision-making and provides high explainability by modeling the relationship between medical images and clinical records.

In this paper, we focus on the PersAM framework in the context of digital pathology.
Particularly, we deal with malignant lymphoma as the target disease, whereas the proposed PersAM framework can be applied to other images in similar problem settings.
In digital pathology, whole slide images (WSIs) are used as image data, which are large digital images scanned by a scanner.
The image size of WSIs can be up to 100,000$\times$100,000 pixels and the tissue regions of WSIs have both tumor and normal regions mixedly.
Therefore, it is especially important in digital pathology to identify the attention regions in a vast image and diagnose focusing on some areas in the tissue specimen.
%
%
%
For examples of malignant lymphoma, pathologists diagnose diffuse large B-cell lymphoma (DLBCL) focusing on large cells in a tissue specimen, while they diagnose follicular lymphoma (FL) focusing on follicular structures in a tissue specimen.
%
In practical diagnosis, as mentioned above, pathologists observe such regions considering a patient's clinical record information that includes basic profiles and results of some examinations.
%
As the main target problem, we are concerned in this paper with the PersAM framework in a digital cancer pathology task, where clinical records can be used together with the WSIs of tissue specimens as patient-specific information.

The problem of attention region estimation in digital cancer pathology can be formulated as a weakly supervised learning problem because only the class label for the entire image is given---the annotations for the attention region are not.
Some public database has pathologists' annotations for tumor regions, but most problem settings that employ other private datasets have no patch-level annotations and only patient-level annotations.
Hence, in digital pathology using WSIs, attention region estimation and each machine learning task should be generally performed with only patient-level annotations.
\emph{Multiple Instance Learning (MIL)}~\cite{ilse2018attention,sahasrabudhe2020deep,li2021multi} is one method used for such weakly supervised attention region estimation problems.
In MIL, an image patch is considered an \emph{instance} and the entire image (or set of a large number of patches) is considered a \emph{bag}.
The problem then reduces to estimating the label of each image patch given the label of the bag (e.g., tumor or normal), where the image patches estimated to be tumors are interpreted as the attention regions.
In the context of MIL in digital pathology, attention-based MIL is well-known as a successful method.~\cite{ilse2018attention,hashimoto2020multi,li2021dual}.
An attention-based MIL can compute attention weights that indicate how each instance contributes to the classification result.
Instances that have higher attention weights in WSI are interpreted as tumor regions in attention-based MIL for digital pathology.
In this study, we would like to introduce the PersAM framework that can adaptively change attention regions depending on different clinical record information even if input WSI is the same.
In the case of the aforementioned attention-based MIL, there is no mechanism to change the attention regions when different clinical record information is input to the same WSI since attention weights are calculated independently for each instance without considering the relationship between medical images and clinical records.
The proposed PersAM method employs a variant of the Transformer architecture to encode the relationships between the medical images and corresponding clinical records.
In the Transformer architecture, the relationships between multiple components are expressed in the form of \emph{attentions}~\cite{vaswani2017attention}.
The application of Transformer architecture to computer vision can calculate attention regions calculated by encoding the relationship between image patches~\cite{dosovitskiy2020image}.
The Transformer architecture could be expanded even into multimodal inputs such as images and table data, which encodes the relationship between each instance of multimodal inputs~\cite{yu2019multimodal,chen2020uniter}.
By combining the medical images and clinical records and then computing the attentions, the proposed method enables \emph{personalized attention}, which represents the strength of the relationship between each clinical record and each region (patch) in the image.

\begin{figure}[t!]
\begin{center}
\includegraphics[width=1.0\linewidth]{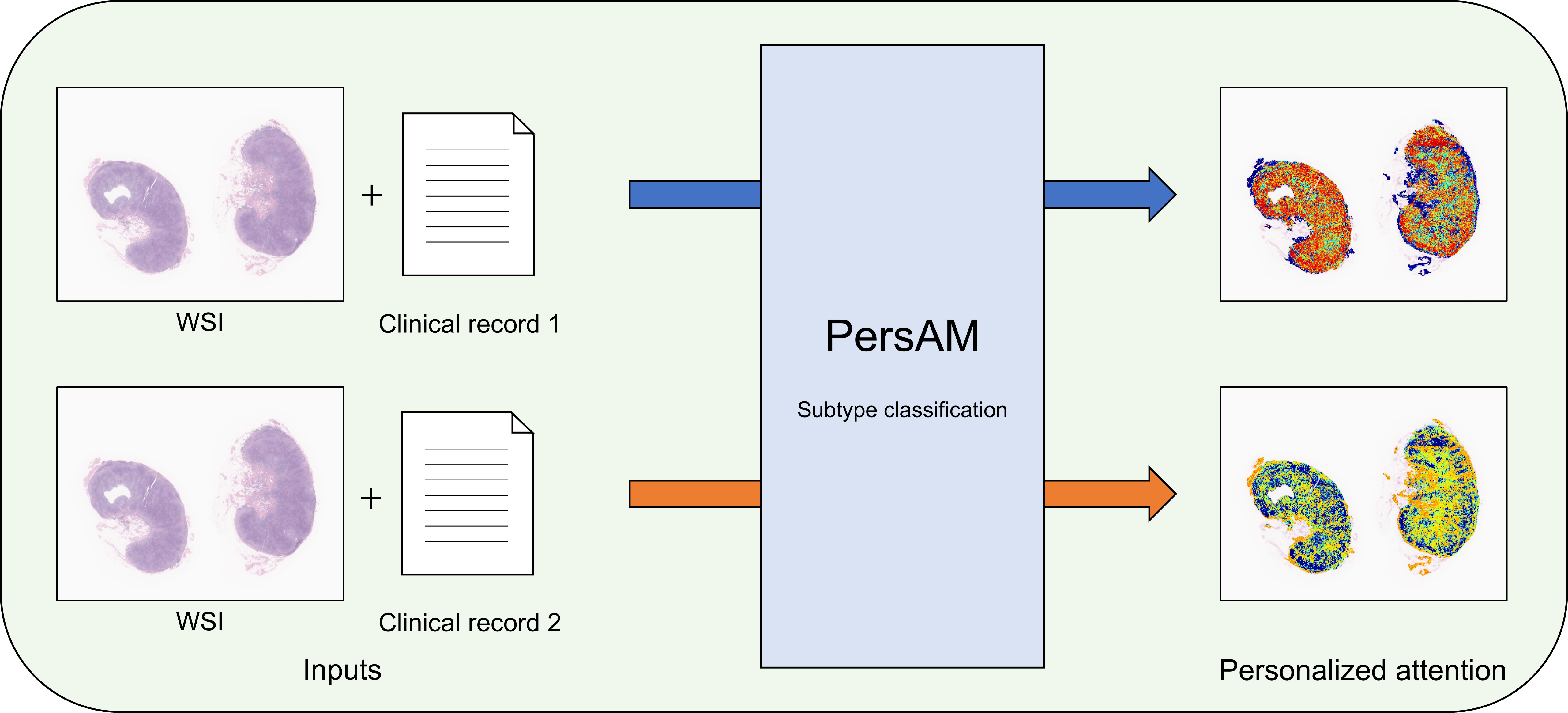}
\end{center}
\caption{
Overview of the proposed PersAM method.
A WSI and a clinical record are fed into the model together.
Regions with red color in Personalized attention represent attention regions in each output.
The PersAM method provides us the personalized attention according to the clinical record, where attentions change depending on clinical factors, even if the same WSI is input.
}
\label{fig:concept}
\end{figure}

In this study, we present a weakly supervised attention region estimation problem formulated as an MIL problem and propose the \emph{PersAM} method to obtain attention regions that can be adaptively changed according to patient clinical records.
Figure~\ref{fig:concept} illustrates the concept of the proposed PersAM method, where different clinical record information is given to the same WSI as inputs.
Regions with red color in \emph{Personalized attention} represent attention regions in each output, and the PersAM method can provide different attention regions depending on different clinical records even if input WSIs are the same.
This mimics a pathologist's decision-making where he/she observes tumor-specific regions in the tissue specimen considering the corresponding patient's clinical record.
With a slight abuse of terminologies, we refer to both the framework and our proposed method as \emph{PersAM} in this work.
The proposed PersAM method enables us to provide two types of personalized attentions: \emph{exploratory} and \emph{explanatory attentions}.
Exploratory attention is a class-independent attention that is determined solely by a WSI and a clinical record, i.e., the first regions of interest to the pathologist when observing a tissue specimen.
On the other hand, explanatory attention is a class-dependent attention that is determined by a WSI, a clinical record, and class information, i.e., the regions of interest to the pathologist when predicting a disease.
To obtain these attentions, the proposed model has a Transformer architecture that can encode the relationship among images, clinical records, and class information.

To demonstrate the effectiveness of the proposed PersAM method, we applied it to the pathological subtype classification of 842 patients with malignant lymphoma.
The training dataset consisted of WSIs and clinical records, where each WSI was a gigapixel image of an entire pathological tissue slide: the clinical record included the age, gender, target organ of the tissue section, interview with a doctor, and blood test results.
By combining pathological images and clinical records, the proposed method performed better than several baseline methods.
Furthermore, we confirmed that the proposed PersAM method can successfully provide personalized attention in the Transformer architecture.

The main contributions of this work are summarized as follows.
\begin{itemize}
\item First, inspired by medical image diagnoses by pathologists in clinical practice, we introduce a framework for a personalized attention mechanism in which the attention is determined on the basis of patient-specific information.
\item Second, for the problem of weakly supervised attention region estimation based on MIL, we propose a variant of the Transformer architecture.
\item Third, we apply the proposed model to a large-scale digital pathology task to demonstrate the effectiveness of the proposed framework and method.
\end{itemize}

\section{Preliminaries}
\subsection{Problem setup}
In this paper, we focus on the PersAM framework in the context of subtype classification for digital cancer pathology.
Let $[N] = \{1, \ldots, N\}$ be the set of natural numbers up to $N$.
The training dataset is denoted as $\{(\mathbb{X}_{n}, \mathbb{T}_{n}, \mathbb{Y}_{n})\}_{n \in [N]}$, where $N$ is the number of patients and each of $\mathbb{X}_{n}$, $\mathbb{T}_{n}$, and $\mathbb{Y}_{n}$ represents the pathological image, clinical record, and subtype class label of the $n^{\rm th}$ patient, respectively.
The image $\mathbb X_n$ is a digitally scanned WSI of the entire pathological specimen.
Because the WSI $\mathbb X_n$ is usually a huge image of gigapixel size, it is too large to be directly fed into the model.
Therefore, image patches extracted from $\mathbb X_n$ are used as the inputs to the model, and we write $\mathbb X_n = \{\bm x_{n,\ell}\}_{\ell \in [L_n]}$ where $\bm x_{n,\ell}$ is the $\ell^{\rm th}$ image patch taken from the $n^{\rm th}$ patient and $L_n$ is the number of image patches taken from $\mathbb X_n$.
The clinical record $\mathbb T_n$ is represented as a set of numerical vectors and denoted as $\mathbb T_n = \{\bm t_{n,m}\}_{m \in [M]}$, where $\bm t_{n,m}$ is the $m^{\rm th}$ clinical factor represented as a vector and $M$ is the number of clinical factors.
For example, in \S4, we consider the case with $M=2$ where the first clinical factor is the patient profile (such as age and gender) and the results of a medical interview, whereas the second clinical factor is a set of blood test results.
In our clinical records, the patient profiles are represented by integer values or binary labels, the results of a medical interview are represented by binary labels, and the blood test results are represented by continuous values, respectively.
If a clinical record has text information as findings, it can be used as a clinical factor $\bm t$ after vectorizing in some manner.
The details of clinical record information is explained in \S4.
The subtype class label $\mathbb Y_n$ is represented as a $C$-dimensional one-hot vector.

In digital cancer pathology, a WSI includes both tumor cells and normal cells, and subtype diagnosis is conducted on the basis of a subset of the tumor cells.
This means that, among the image patches taken from a WSI, only some of them are considered to contain useful information for subtype classification.
We regard that image patch subset as the attention region.
We represent the attention degree of each image patch as an attention weight.
Given a pathological WSI and a clinical record, our model provides the attention weights---which each represents the importance of each image patch---and then makes a subtype classification based on the attention weights.
As an example of the application of the PersAM framework in the digital pathology problem, we consider the case where the attention weights vary according to the clinical records.
Clinical records possibly contain two types of information: i) the parts of the pathological specimen that should be observed and ii) which subtype it is likely to be classified under.
In this study, we consider two types of personalized attentions, called \emph{exploratory} and \emph{explanatory attentions}, each of which is respectively obtained from each of these two types of information.
We introduce a variant of the Transformer architecture that can provide both the exploratory attention weight and the explanatory attention weight of each image patch.

%
%
%
%
%
%

\subsection{Related works}

\paragraph{Digital pathology}
Pathological diagnosis plays an important role in medicine.
Various computer-aided diagnosis methods for pathological images have been developed for various problems, such as classification~\cite{hou2016patch,mousavi2015automated,hashimoto2020multi},
tumor region identification~\cite{cirecsan2013mitosis,cruz2014automatic,bejnordi2017diagnostic}, segmentation~\cite{xu2015deep,naylor2018segmentation,tokunaga2019adaptive,tanizaki2020computing},
survival prediction~\cite{zhu2017wsisa,wulczyn2020deep,huang2021integration}, and similar image retrieval~\cite{hegde2019similar,kalra2020yottixel,hashimoto2021case}.
In digital pathology, a digital scan of the entire pathology specimen, called a WSI, is used as the target image.
Because a WSI is usually huge (e.g., 100,000$\times$100,000 pixels), it cannot be directly fed into a model.
Therefore, image patches extracted from the WSI are often used as the inputs to a model.
In pathological diagnosis based on WSIs, it is important to note that WSIs contain both tumor cells and normal cells.
Therefore, if there is no annotation of the tumor cell region, it is necessary to first identify the tumor cell region and then make a pathological diagnosis.
This problem is a weakly supervised learning problem in the sense that only the WSI is labeled and the tumor cell region is not.

\paragraph{Multiple instance learning (MIL)}
MIL is a weakly supervised learning problem in which labels are not given for instances but for a group of instances called a bag.
In an MIL formulation of a binary classification problem, it is assumed that a positive bag contains at least one positive instance, whereas a negative bag contains only negative instances.
By considering a WSI as a bag and an image patch as an instance, the subtype classification problem can be interpreted as an MIL problem in which class-specific image patches (e.g., tumor patches) are considered positive instances.
Several MIL approaches have been developed for digital pathology tasks~\cite{ilse2018attention,sahasrabudhe2020deep,li2021multi}.
Among them, attention-based MIL~\cite{ilse2018attention,hashimoto2020multi,li2021dual} is particularly useful because the identified attention regions can be interpreted as class-specific image patches.

\paragraph{Attention
 and explanation}
Although the development of deep learning techniques has dramatically improved the accuracies of many medical image analysis tasks, it is critically important for medical practice to develop techniques that provide 
explanation of the results.
Various visualization methods, such as Grad-CAM~\cite{selvaraju2017grad}, have been proposed to interpret and explain the rationale for classification results.
Singla et al. \cite{singla2021using} proposed a method that visualizes the medical image regions that serve as the basis for the classification of diseases for each concept derived from clinical report analysis.
However, most visualization methods visualize the regions that contribute to the classification results after the classifier is applied to given images (rather than using clinical reports for finding and visualizing the image regions that contribute to the classification results, as in our proposed method).
The attention-based MIL described above can also be considered as a method to provide 
explanation of the results because attention can be visualized as an informative region for making decisions.
In this study, we employ the Transformer architecture~\cite{vaswani2017attention} as a basis for estimating the attention region in medical images.
Although the Transformer was originally developed for natural language processing (NLP) tasks, it has been demonstrated to be effective for general computer vision tasks~\cite{dosovitskiy2020image}, including medical image analysis~\cite{rymarczyk2021kernel,shao2021transmil,lu2021smile,gao2021instance}.
In particular, the Transformer has been effectively used to aggregate bag features in an MIL setting~\cite{shao2021transmil}.
The Transformer architecture can encode the relationship among a pair of components in input data, e.g., between two words in the case of NLP tasks, and between two image patches in the case of computer vision tasks.
This study was inspired by the use of the Transformer in the context of vision and language~\cite{yu2019multimodal,chen2020uniter}, where it has been demonstrated that a change in the language token can change the attention of the image token.
The Transformer has a mechanism to quantify the relevance of multimodal information in the form of attentions.
The encoding mechanism for the relationship of input data can be expanded into multimodal input including image patches and table data, which enables computing attention from an image patch to table data, or attention from table data to an image patch.
We expect that encoding the relationship between image patches and clinical factors can provide more appropriate personalized attentions based on the clinical record.

\paragraph{Multimodal learning}
In clinical practice, doctors obtain additional information from patients' clinical records and image diagnoses are performed by considering such clinical factors as prior information.
In multimodal analysis, clinical records that include the basic information of patients and some examination results can often be used in addition to digital images~\cite{yala2019deep,yap2018multimodal,thung2017multi,nie2019multi}.
Yala et al. \cite{yala2019deep} used a combination of mammography images and patient data (basic information, medical history, etc.) and demonstrated that the performance of breast cancer risk prediction could be improved.
Multimodal analysis of dermoscopic images and patient data (age, gender, and body location) has also been studied to enhance the accuracy of skin lesions classification and melanoma detection~\cite{yap2018multimodal}.

Multimodal analyses of medical images and clinical records have primarily been performed on radiology images, but recent works have reported that pathological images can also be combined with clinical records to improve the task performance.
Li et al. \cite{li2021multi} combined tabular clinical data with histological images in an MIL setting, where eighteen attributes, including age, genes, and tumor location, were used as inputs with multiscale histological images.
Additional clinical factors have also been used in a mixture-of-experts model as inputs of a gating network~\cite{sahasrabudhe2020deep}.
Chen et al. \cite{chen2021multimodal} proposed a Transformer-based multimodal model for survival prediction using images and genetic data. Their method could visualize co-attentions between images and genetic information.
However, these previous works primarily focused on performance improvement by using multimodal inputs, and detailed effects on attention regions by additional clinical factors have not been reported.
The advantage of the proposed PersAM method is that it can provide personalized attention regions according to the clinical records, which mimics the actual pathological practice of human expert pathologists.

\section{Proposed Method}

This study was conducted to develop an AI system for pathological diagnosis that mimics the actual diagnosis process of human pathologists.
When a pathologist makes a diagnosis, they have patient information based on the clinical record, which is used as prior knowledge of the parts of the pathological image on which to focus.
We call such an attention region in the early exploratory phase of the diagnosis \emph{exploratory attention}.
Exploratory attention is a class-independent attention that is determined solely by a WSI and a clinical record.
Furthermore, after a pathologist has made a diagnosis, they should be able to explain which part of the pathological image they focused on.
We call such an attention region in the later explanatory phase of the diagnosis \emph{explanatory attention}.
Explanatory attention is a class-dependent attention that is determined by a WSI, a clinical record, and class information.

Given the WSI and the clinical record of a patient, the proposed PersAM method can identify both exploratory and explanatory attention regions and make a diagnosis based on those identified attention regions.
In \S3.1, we first introduce a Transformer-based network structure that enables pathological diagnosis based on the two types of attentions.
Then, in \S3.2, we describe how the network can be used to identify both exploratory and explanatory attention regions to make a pathological diagnosis.
The key property of the proposed PersAM method is that both exploratory and explanatory attentions can be adaptively changed according to patient clinical records even if the same WSI is given to the model.
This mimics the actual diagnostic process employed by human pathologists, making the AI system highly 
explainable.

In this section, for ease of notation, when there is no ambiguity, we omit the subscript $n$ in referring to the $n^{\textrm{th}}$ patient.\\

\subsection{Proposed network structure}

\begin{figure}[t!]
\begin{center}
\includegraphics[width=1.0\linewidth]{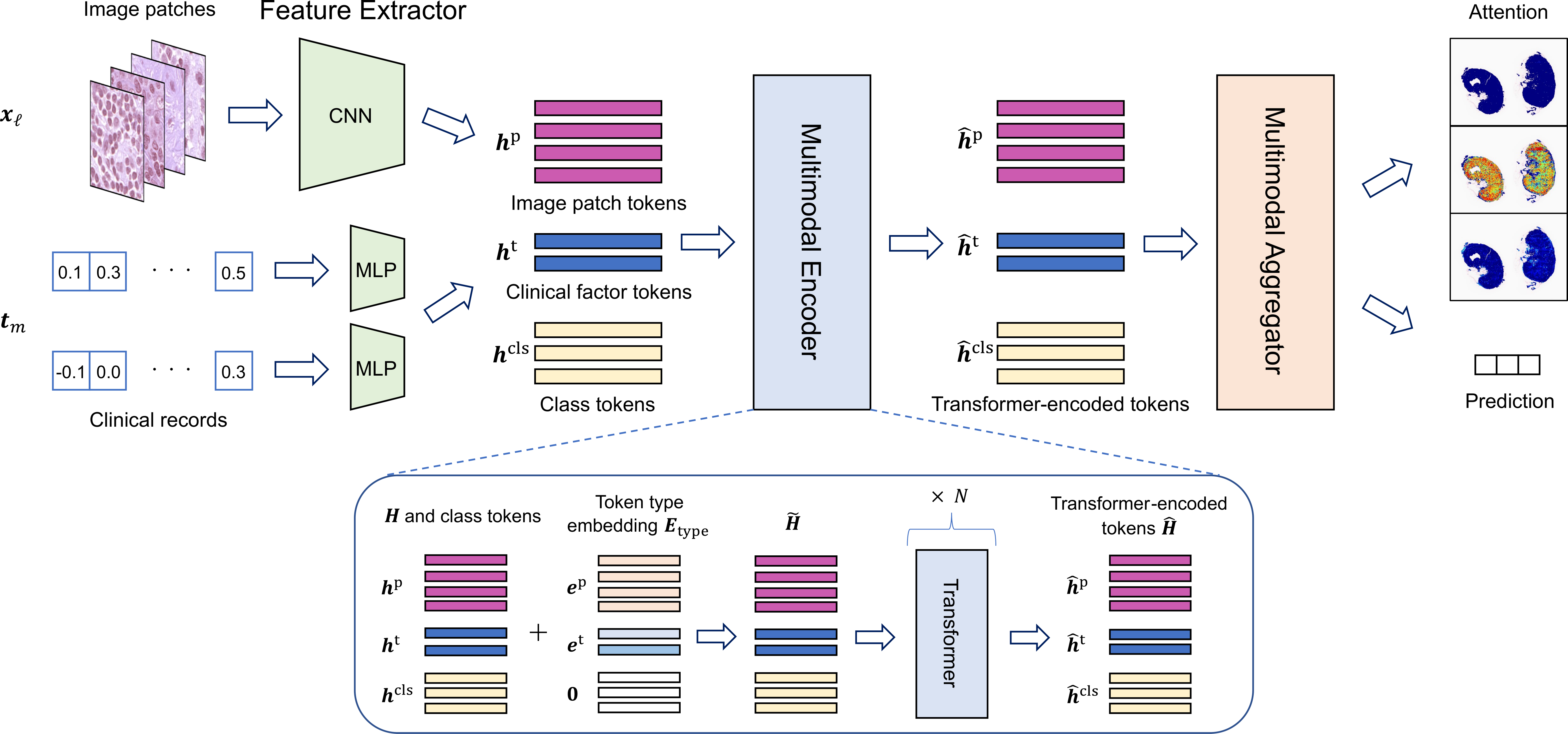}
\end{center}
 \caption{Illustration of the proposed network structure. The network consists of three components: \emph{feature extractor}, \emph{multimodal encoder}, and \emph{multimodal aggregator}.
 The feature extractors compute feature vectors $\boldsymbol{H}$ for each of the image patches and clinical factors to be fed into the Transformer architecture.
The multimodal encoder has a role to characterize the relationship among multimodal information consisting of image patches, clinical factors, and class information.
The multimodal aggregator aggregates the Transformer-encoded tokens for obtaining exploratory/explanatory attentions and subtype classification results.
 }
\label{fig:method}
\end{figure}

To realize pathological diagnosis based on exploratory and explanatory attentions, as shown in Fig.~\ref{fig:method}, we propose a new network structure consisting of three components: \emph{feature extractor}, \emph{multimodal encoder}, and \emph{multimodal aggregator}.
We describe each of these three components below.

\subsubsection{Feature extractor}

In this section, to simplify the notations, we describe the MIL setting where the entire WSI is considered as a \emph{bag} and each of the multiple patches taken from the WSI as an \emph{instance}.
In \S4, we consider the MIL setting where a WSI contains multiple bags. See \S4 for details.
Let $\mathbb X = \{\bm x_\ell\}_{\ell \in [L]}$ be the WSI and $\mathbb T = \{\bm t_m\}_{m \in [M]}$ be the clinical record of a patient, where $\bm x_\ell, \ell \in [L], $ is the $\ell^{\rm th}$ patch image taken from the WSI $\mathbb X$, whereas $\bm t_m, m \in [M], $ is the $m^{\rm th}$ clinical factor.
The role of the feature extractor is to compute a feature vector for each of the image patches and clinical factors so that input data can be used in the Transformer architecture.
For image patches, we employ a convolutional neural network (CNN) $f$: $\boldsymbol{x}_{\ell} \mapsto \boldsymbol{h}_{\ell}^{\mathrm{p}}$ that maps an image patch $\boldsymbol{x}_{\ell}$ to a feature vector $\bm h_{\ell}^{\mathrm{p}} \in \mathbb{R}^R$, where $R$ is the dimension of the feature vector.
For clinical factors, we employ a simple multi-layer perceptron (MLP) $g_{m}$: $\boldsymbol{t}_{m} \mapsto \boldsymbol{h}_m^{\mathrm{t}}$ that maps the vector of the $m^{\rm th}$ clinical factor into a feature vector $\boldsymbol{h}_{m}^{\mathrm{t}} \in \mathbb{R}^R$ that has the same dimension as $\bm h_{\ell}^{\mathrm{p}}$.
We denote the sets of trainable parameters for $f$ and $\{g_m\}_{m \in [M]}$ as $\theta_{\mathrm{f}}$ and $\theta_{\mathrm{g}}$, respectively.
We denote the combined feature vectors as
\begin{align}
\boldsymbol{H} = \{\boldsymbol{h}_{\ell}^{\mathrm{p}}\}_{\ell \in [L]} \cup \{\boldsymbol{h}_{m}^{\mathrm{t}}\}_{m \in [M]}.
\end{align}

\subsubsection{Multimodal encoder}
The multimodal encoder characterizes the relationship between multimodal information.
We implement this component using a \emph{Transformer}.
The Transformer was initially developed for NLP tasks, where each feature vector is called a \emph{token}.
In the proposed network structure, in addition to the image patch tokens $\{\bm h_\ell^{\rm p}\}_{\ell \in [L]}$ and clinical factor tokens $\{\bm h_m^{\rm t}\}_{m \in [M]}$, we introduce \emph{class tokens} $\{\bm h_c^{\rm cls}\}_{c \in [C]}$, which are considered trainable parameters.
Let
\begin{align}
\tilde{\boldsymbol{H}} = [\boldsymbol{h}_{1}^{\mathrm{p}}, \dots, \boldsymbol{h}_{L}^{\mathrm{p}}, \boldsymbol{h}_{1}^{\mathrm{t}} ,\dots,\boldsymbol{h}_{M}^{\mathrm{t}},
\boldsymbol{h}^{\mathrm{cls}}_{1}, \dots, \boldsymbol{h}^{\mathrm{cls}}_{C}]^{\top} + \boldsymbol{E}_{\mathrm{type}},
\end{align}
where $\boldsymbol{E}_{\mathrm{type}} \in \mathbb{R}^{R \times (L+M+C)}$ is called \emph{token type embedding} and is used to characterize the type of tokens.
Here, the token type embedding is defined as
\begin{align}
\boldsymbol{E}_{\mathrm{type}} = [\overbrace{\boldsymbol{e}^{\mathrm{p}}, \dots, \boldsymbol{e}^{\mathrm{p}}}^{L}, \boldsymbol{e}^{\mathrm{t}}_{1}, \dots, \boldsymbol{e}^{\mathrm{t}}_{M},
\overbrace{\boldsymbol{0}, \dots, \boldsymbol{0}}^{C}]^{\top},
\end{align}
where $\boldsymbol{0}$ denotes a $C$-dimensional zero vector.
Each token in $\tilde{\boldsymbol{H}}$ is characterized as either of image patch, clinical factor, or class token by this token embedding.
Note that the same type token $\boldsymbol{e}^{\mathrm{p}}$ is used for all $L$ image patches because the image patches are randomly sampled from WSI.
Token type embedding parameters $\bm e^{\rm p}$ and $\{\bm e_m^{\rm t}\}_{m \in [M]}$ are considered trainable parameters.

We denote the Transformer as a function $\mathcal T: \tilde{\bm H} \mapsto \hat{\bm H}$, where $\hat{\bm H}$ is the collection of the outputs of the Transformer called \emph{Transformer-encoded tokens}, denoted as
\begin{align}
 \hat{\bm H} = \{
 \{\hat{\bm h}_c^{\rm cls}\}_{c \in [C]},
 \{\hat{\bm h}_\ell^{\rm p}\}_{\ell \in [L]},
 \{\hat{\bm h}_m^{\rm t}\}_{m \in [M]}
 \}.
\end{align}
We denote the set of trainable parameters for the Transformer as
\begin{align}
  \theta_{\rm enc} = \{\bm h_1^{\rm cls}, \ldots, \bm h_C^{\rm cls}, \bm e^{\rm p}, \bm e_1^{\rm t}, \ldots, \bm e_M^{\rm t}, \theta_{\rm tf}\},
  \end{align}
where
$\theta_{\rm tf}$ is the other general parameters in the Transformer.
By feeding tokens into the Transformer encoder several times repeatedly, $\hat{\bm H}$ can encode the relationship among image patches, clinical factors, and class tokens.
Here each element of the self-attention map for $\hat{\bm H}$ corresponds to the relationship between two tokens, and represents how a token is focused when the other token is given together as input data.

\subsubsection{Multimodal aggregator}
The multimodal aggregator aggregates the Transformer-encoded tokens for obtaining exploratory/explanatory attentions and subtype classification results.
Let
\begin{align}
&
\boldsymbol{q}^{\mathrm{cls}}_{c} = \boldsymbol{W}_{\mathrm{q}}\hat{\boldsymbol{h}}^{\mathrm{cls}}_{c}, \quad c \in [C], \\
&
\boldsymbol{k}^{\mathrm{p}}_{\ell} = \boldsymbol{W}_{\mathrm{k}}\hat{\boldsymbol{h}}^{\mathrm{p}}_{\ell},
~
\boldsymbol{v}^{\mathrm{p}}_{\ell} = \boldsymbol{W}_{\mathrm{v}}\hat{\boldsymbol{h}}^{\mathrm{p}}_{\ell}, \quad \ell \in [L], \\
&
\boldsymbol{q}^{\mathrm{t}}_{m} = \boldsymbol{W}_{\mathrm{q}}\hat{\boldsymbol{h}}^{\mathrm{t}}_{m},
~
 \boldsymbol{k}^{\mathrm{t}}_{m} = \boldsymbol{W}_{\mathrm{k}}\hat{\boldsymbol{h}}^{\mathrm{t}}_{m}, \quad m \in [M],
\end{align}
where $\{\bm q_c^{\rm cls}\}_{c \in [C]}$ and $\{\bm q_m^{\rm t}\}_{m \in [M]}$ are called \emph{queries} for class tokens and clinical factor tokens, respectively; $\{\bm k_\ell^{\rm p}\}_{\ell \in [L]}$ and $\{\bm k_m^{\rm t}\}_{m \in [M]}$ are called \emph{keys} for image patch tokens and clinical factor tokens, respectively; and $\{\bm v_\ell^{\rm p}\}_{\ell \in [L]}$ is called \emph{values} for image patch tokens in the context of the Transformer.
The matrices $\boldsymbol{W}_{\mathrm{q}}, \boldsymbol{W}_{\mathrm{v}}, \boldsymbol{W}_{\mathrm{k}} \in \mathbb{R}^{R \times R}$ are trainable parameters.
We denote these three matrices collectively as $\theta_{\rm agg} = \{\bm W_{\mathrm{q}}, \bm W_{\mathrm{v}}, \bm W_{\mathrm{k}}\}$.

The inner product between a query and a key represents the relevance between the corresponding multimodal information.
First, the relevance between each image patch $\bm x_\ell$ and each class $c$ is written as
\begin{align}
 \label{eq:relevance1}
 a_{\ell, c} = \sigma(\bm q_c^{{\rm cls} \top} \bm k_\ell^{\rm p}), ~ (\ell, c) \in [L] \times [C],
\end{align}
where $\sigma(\cdot)$ is the sigmoid function.
Next, the relevance between each image patch $\bm x_\ell$ and the set of $M$ clinical factors $\{\bm t_m\}_{m \in [M]}$ is written as
\begin{align}
 \label{eq:relevance2}
 \psi_\ell = \frac{1}{M} \sum_{m \in [M]} \sigma(\bm q_m^{{\rm t} \top} \bm k_\ell^{\rm p}), ~ \ell \in [L].
\end{align}
Furthermore, the relevance between each class and the set of $M$ clinical factors $\{\bm t_m\}_{m \in [M]}$ is written as
\begin{align}
 \label{eq:relevance3}
 \phi_c = \frac{1}{M} \sum_{m \in [M]} \sigma(\bm q_c^{{\rm cls} \top} \bm k_m^{\rm t}), ~ c \in [C].
\end{align}
The three types of relevance information in \eq{eq:relevance1}--\eq{eq:relevance3} are used to obtain exploratory/explanatory attentions and subtype classification results.

\subsection{Exploratory/Explanatory attentions and subtype classifications}
\begin{figure}[t!]
\begin{center}
\includegraphics[width=1.0\linewidth]{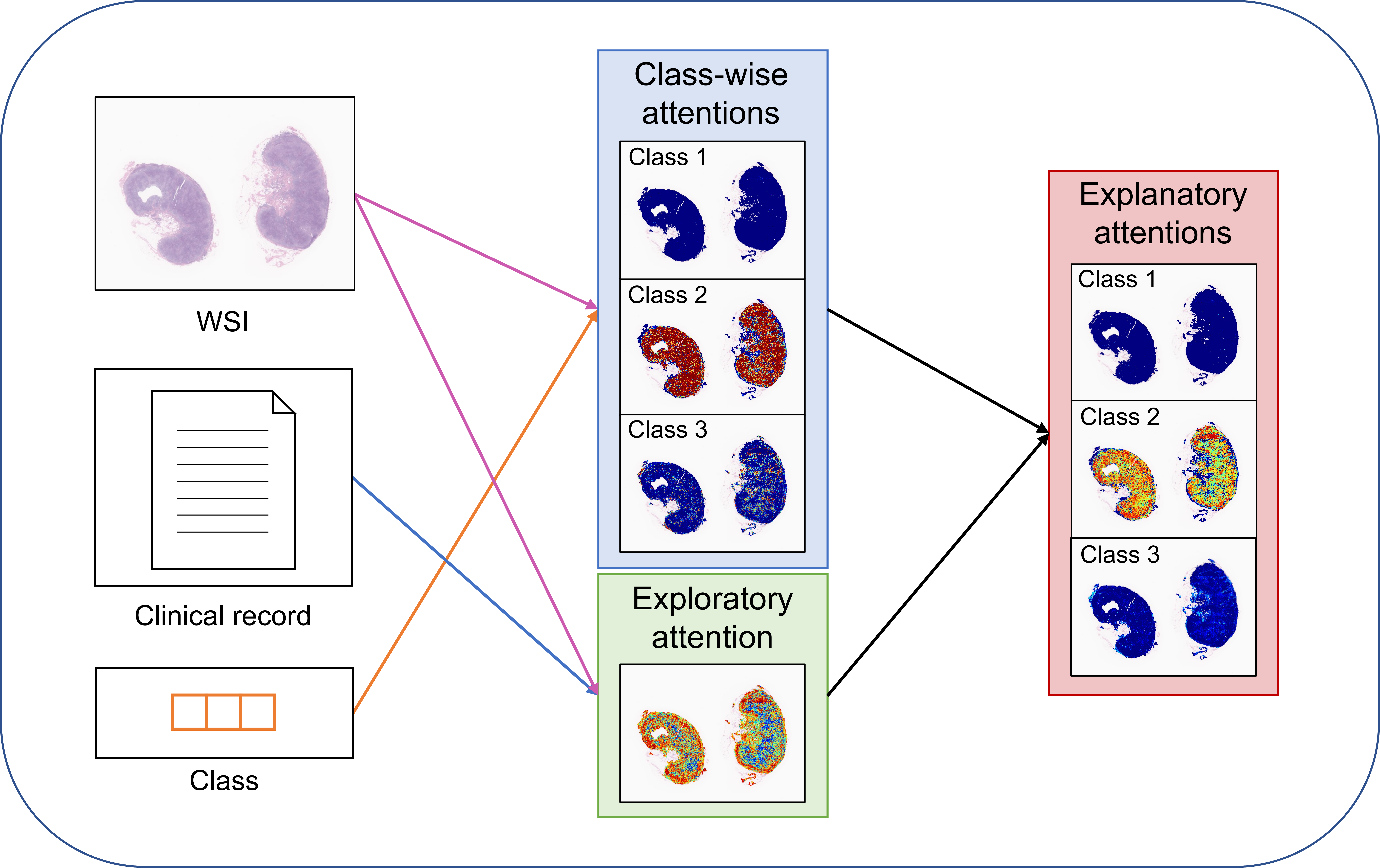}
\end{center}
 \caption{
The three types of attentions considered in the proposed method.
 The class-wise attentions are obtained based on the relevance between image patches and class tokens.
 For this example, the model focuses on the almost entire WSI to identify the case as class 2 when only the WSI is used for the class prediction.
 The exploratory attentions are obtained based on the relevance between image patches and clinical factors and given to the regions focused on regardless of which class the WSI belongs to.
 The explanatory attentions are obtained by filtering the class-wise attentions with the exploratory attentions and are provided as a reason for the final determination by considering the relationship among multimodal information.
}
\label{fig:aggregator_simple}
\end{figure}

Based on the relevance information in \eq{eq:relevance1}--\eq{eq:relevance3}, we obtain three types of attentions: \emph{class-wise attentions}, exploratory attentions, and explanatory attentions.
Fig.~\ref{fig:aggregator_simple} illustrates the three types of attentions.
We call $\{a_{\ell, c}\}_{(\ell, c) \in [L] \times [C]}$ class-wise attentions because they are obtained as the relevance between the $\ell^{\rm th}$ image patch and the $c^{\rm th}$ class token without clinical factors.
We regard $\{\psi_{\ell}\}_{\ell \in [L]}$ as exploratory attentions because they are obtained as the relevance between the $\ell^{\rm th}$ image patch and the set of clinical factors $\{\bm t_m\}_{m \in [M]}$.
Note that the exploratory attentions are class-independent; thus, they can be considered as the attention regions in the WSI in the early exploratory phase of the diagnosis.
The explanatory attentions are obtained by combining the class-wise attentions and the exploratory attentions as follows:
\begin{align} 
 a^\prime_{\ell, c} = a_{\ell, c} \phi_c \psi_\ell, ~ (\ell, c) \in [L] \times [C].
\end{align}
It can be interpreted that the explanatory attentions are obtained by filtering the class-wise attentions with the exploratory attentions.

The subtype classification results are obtained based on a linear combination of the aggregate feature vector
\begin{align}
 \bm z = \sum_{\ell \in [L]} \frac{a_\ell^\prime}{\sum_{\rho=1}^L a_\rho^\prime} \bm v_\ell^{\rm p},
\end{align}
where $a_\ell^\prime = {\rm max}(a_{\ell,1}^\prime, \ldots, a_{\ell,C}^\prime), \ell \in [L]$.
Then, the class-wise probabilities are obtained by using a neural network (NN) with the softmax operator $g_{\rm clf}: \bm z \mapsto \hat{\mathbb Y}$ as follows:
\begin{align}
 \hat{\mathbb Y} = g_{\rm clf}(\bm z; \theta_{\rm clf}),
\end{align}
where
$\hat{\mathbb Y}$
is a $C$-dimensional vector whose $c^{\rm th}$ element represents the probability that the subtype of the patient is class $c$ and $\theta_{\rm clf}$ is the set of trainable parameters.

When the network is trained, the loss function consists of two loss components.
The first loss component is simply the cross-entropy loss $\cL_{\rm ce}(\mathbb Y, \hat{\mathbb Y})$ between the true one-hot class vector $\mathbb Y$ and the predicted class probability vector $\hat{\mathbb Y}$.
The second loss component is considered to take into account the specific property of the MIL setting, where the bag (WSI) is positive if any of the instances (image patches) is positive.
To formulate this specific property, we consider
\begin{align}
 \label{eq:pi-c}
 \pi_c = 1 - \prod_{\ell \in [L]}(1 - a_{\ell, c}^\prime), ~ c \in [C],
\end{align}
where $\pi_{c}$ is close to one if there exists at least one image patch with an attention value $a_{\ell, c}^\prime$ close to one.
The second loss component is defined as the binary cross-entropy $\cL_{\rm bce}(\mathbb Y_{c}, \pi_{c})$ between $\mathbb Y_{c}$, the $c^{\rm th}$ element $\mathbb Y$, and $\pi_c$.
This loss function is inspired by the probability aggregation approach studied in \cite{li2018thoracic}.

\begin{figure}[t!]
\begin{center}
\includegraphics[width=1.0\linewidth]{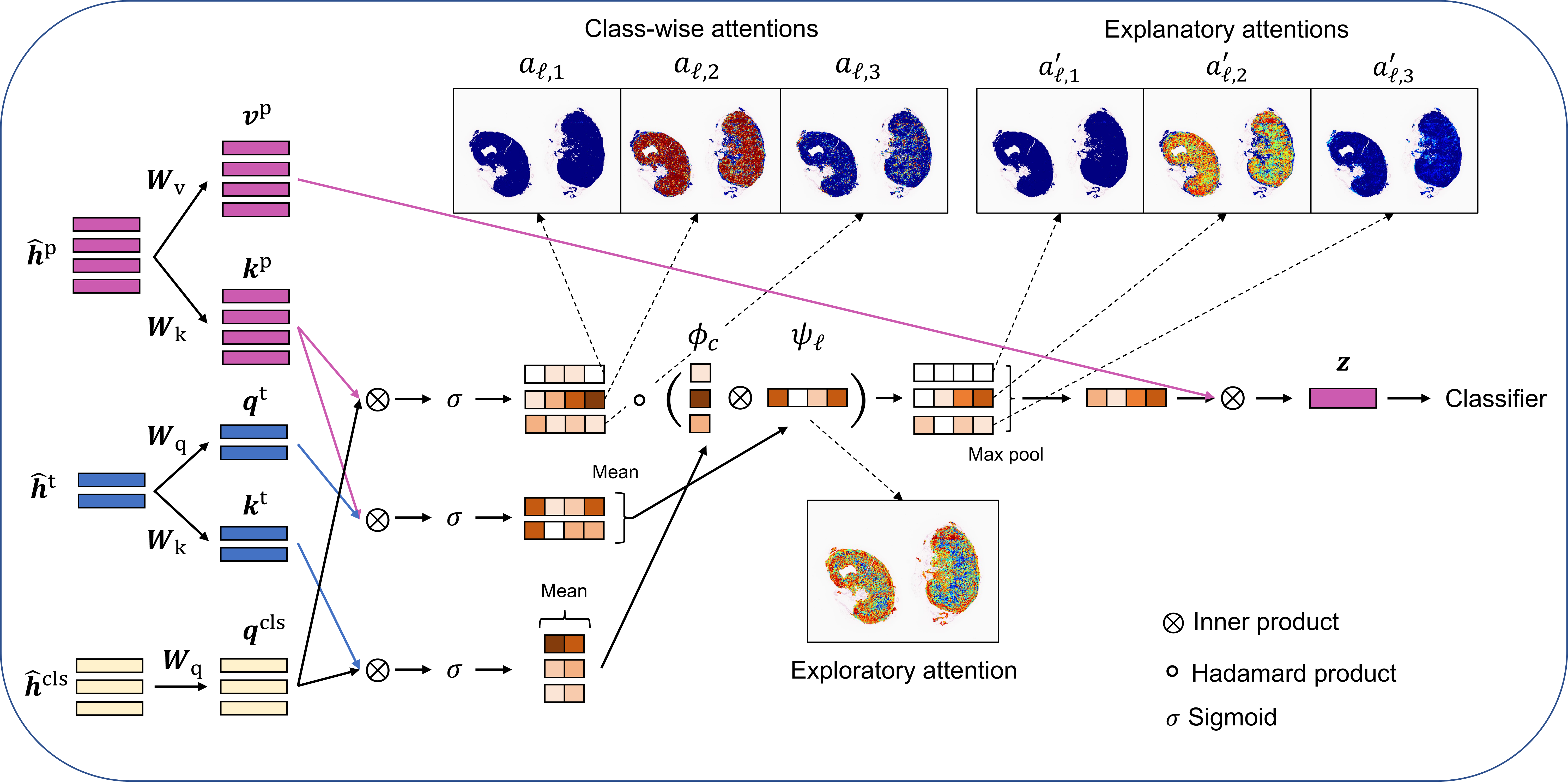}
\end{center}
 \caption{
 Operations performed in the proposed network structure.
 Given a set of image patches and a set of clinical factors, the network predicts the subtype based on exploratory attentions and explanatory attentions.
Class-wise and explanatory attentions are computed for each class, whereas exploratory attention is calculated for a case since it is a class-independent attention based solely on a WSI and a clinical record.
Both exploratory and explanatory attentions can vary depending on clinical records even when an input WSI is the same.
 This mimics the actual pathological diagnosis by human pathologists.
 }
\label{fig:aggregator}
\end{figure}

The proposed network contains trainable parameters $\theta_{\mathrm{f}}$, $\theta_{\mathrm{g}}$, $\theta_{\rm enc}$, $\theta_{\rm agg}$, and $\theta_{\rm clf}$.
All the parameters in the model are simultaneously trained by minimizing the following loss function:
\begin{align}
& L(\theta_{\mathrm{f}}, \theta_{\mathrm{g}}, \theta_{\rm enc}, \theta_{\rm agg}, \theta_{\rm clf}) =  \nonumber \\
& \sum_{n \in [N]} \left\{\cL_{\rm ce}(\mathbb Y_n, \hat{\mathbb Y}_n) + \frac{1}{C} \sum_{c \in [C]} \cL_{\rm bce}(\mathbb Y_{n, c}, \pi_{n, c}) \right\}.
\end{align}
The operations performed in the network structure are illustrated in Fig.~\ref{fig:aggregator}.

\section{Experimental Evaluation}

In the experiments, we first compared the proposed PersAM MIL with several baseline methods to confirm the improvement of classification performance.
Then, the effectiveness of exploratory and explanatory attentions in the proposed method was evaluated.

\subsection{Experimental setting}

\begin{figure}[t!]
\begin{center}
\includegraphics[width=1.0\linewidth]{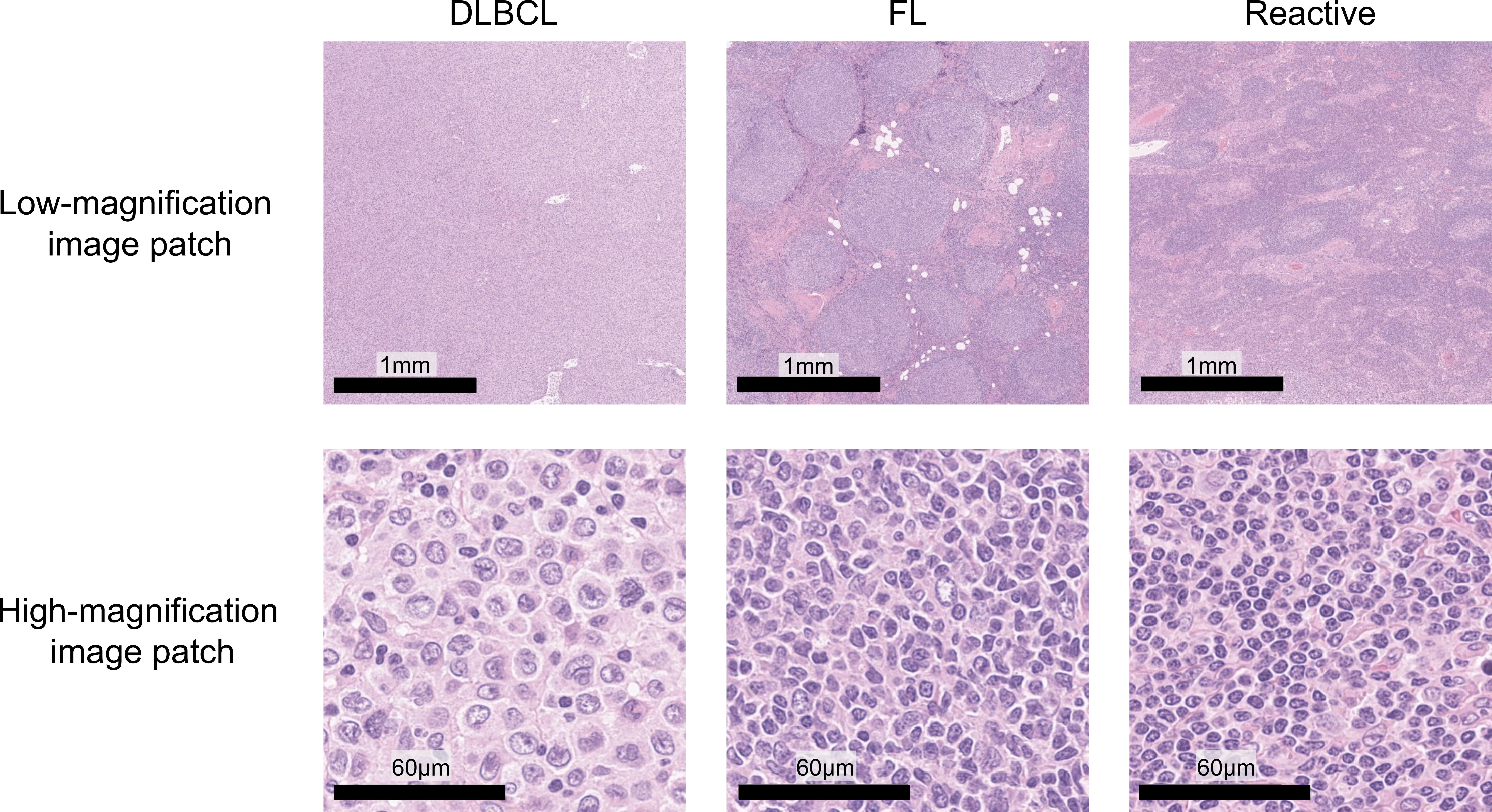}
\end{center}
\caption{
Samples of typical image patches for lymphoma cases.
Each subtype has individual histological features; DLBCL has large tumor cells over a wide regions in the tissue specimen, FL has follicular structures which have tumor cells, and Reactive has diverse cell structures but no tumor cells.%
}
\label{fig:sample}
\end{figure}

\paragraph{Dataset}
Our database of malignant lymphoma was composed of $N=842$ clinical cases with three subtypes: 277 DLBCL, 270 FL, and 295 reactive lymphoid hyperplasia (Reactive).
Figure~\ref{fig:sample} shows sample image patches for typical DLBCL, FL, and Reactive cases.
DLBCL has large tumor cells over a wide region in the tissue specimen, and FL has follicular structures which have tumor cells.
In contrast, Reactive is classified as non-lymphoma, which has diverse cell structures but no tumor cells.
All the patient data were clinically diagnosed by expert hematopathologists and a WSI of a hematoxylin-and-eosin (H\&E)-stained tissue specimen and a clinical record were given for each case.
A gigapixel digitized WSI of the entire H\&E-stained tissue slide, was used as an input image $\mathbb X_n$.
All the glass slides were digitized using a WSI scanner (Aperio GT 450; Leica Biosystems, Germany) at 40x magnification (0.26 {\textmu}m/pixel), where the maximum image size was approximately 100,000$\times$100,000 pixels.
The OpenSlide~\cite{goode2013openslide} software was used for handling WSIs and extracting image patches from $\mathbb X_n$.
An original clinical record includes the definitive subtype $\mathbb Y_n$ and clinical factors $\mathbb T_n$.
Note that we cannot use patch-level annotations, and the class label $\mathbb Y_n$ is given only to a WSI $\mathbb X_n$, not image patches.
The clinical factors $\mathbb T_n$ consist of 28 elements that are summarized by $M=2$ clinical factors: an 18-dimensional vector with patient basic information and interview results and a 10-dimensional vector with blood test results.
The details of items included in each clinical factor are listed in tables~\ref{tab:item1} and \ref{tab:item2}.
%

\begin{table}[t]
 \caption{The detailed items in clinical factor $\bm t_1$ consisting of patient basic information and interview results.
 Other items than age are represented as binary labels, e.g., ``fever'' indicates 1 if a patient had a high fever in the interview.
 }
 \label{tab:item1}
 \centering
 \begin{tabular}{ll}
  \hline
   ~~Item & Value type \\
   \hline \hline
   ~~Age & Nonnegative integer \\
    \hline
   \begin{tabular}{l}
     Gender, organ (lymph node, tonsil, others), fever,\\
     weight loss, hepatomegaly, splenomegaly,\\
     swelling (none, whole body, neck, armpit,\\
     deep abdominal cavity, mouse diameter, septum, others)\\
   \end{tabular}
   & Binary \\
   \hline
 \end{tabular}
\end{table}

\begin{table}[t]
 \caption{The detailed items in clinical factor $\bm t_2$ consisting of blood test results.
All items are represented as continuous values indicating amount or percentile.
 }
 \label{tab:item2}
 \centering
 \begin{tabular}{ll}
  \hline
   Item & Value type \\
   \hline \hline
   RBC, WBC, plt, LDH & Amount \\ \hline
   Stab, seg, eosino, baso, mono, lympho & Percentile \\
   \hline
 \end{tabular}
\end{table}

\paragraph{Implementation details}
In the experiment, $224\times224$-pixel image patches were randomly extracted from an entire WSI and 100 image patches were used as a bag due to the amount of computation and memory.
The corresponding label $\mathbb Y_n$ was assigned to a bag generated from a WSI $\mathbb X_n$, e.g., a bag generated from image patches of a WSI of DLBCL was labeled as DLBCL.
A maximum of 30 bags were generated from a single WSI in our experiment.
The length of the feature vector was set to $R=512$ inspired by TransMIL~\cite{shao2021transmil}.
To obtain feature vector of image patches $\boldsymbol{h}_{\ell}^{\mathrm{p}}$, $f$ employed an ResNet50 \cite{he2016deep} pre-trained with ImageNet and a two-layer NN that had 1024 hidden units, 512 output units, and ReLU as its activation function, where a 2048-dimensional vector after global average pooling layer in ResNet50 was converted into the 512-dimensional feature vector.
Clinical factors were mapped to $\boldsymbol{h}_m^{\mathrm{t}}$ by a two-layer NN that had 256 hidden units, 512 output units, and ReLU as its activation function, where both 18-dimensional and 10-dimensional clinical factors were converted into 512-dimensional feature vectors.
Class tokens $\{\bm h_c^{\rm cls}\}_{c \in [C]}$ were designed as three 512-dimensional vectors, and then we could obtain input data for the Transformer architecture $\tilde{\boldsymbol{H}} \in \mathbb{R}^{512 \times 105}$.
In \eq{eq:pi-c}, to prevent underflow computing, $1 - a_{\ell, c}^\prime$ was normalized to $[0.95, 1.0]$.
The dataset was divided into training, validation, and testing data in the ratio of 3:1:1, and the models were evaluated via five-fold cross-validation where the model that had the smallest validation loss after third epoch was used for testing.

For the setting on Transformer, the number of layers and heads were set to two and eight, respectively, where dropout rate was set to 0.1.
%
%
The classifier $g_{\textrm{clf}}$ was an NN that had a hidden layer with 256 units and an output layer with three units to compute the class probability from a 512-dimensional aggregated feature vector $\bm z$.

For stability in the optimization, label smoothing was applied as a regularization technique in calculating the loss function $\mathcal{L}_{\textrm{bce}}(\mathbb Y_{n,c},\pi_{n,c})$, where the label for a correct class was set to $\mathbb Y_{n,c} = 0.95$ and the labels for incorrect classes were set to $\mathbb Y_{n,c} = 0.05$.
As an optimization method, momentumSGD (nesterov, weight decay=$10^{-4}$) was employed and the training of the model was performed in nine epochs.
Learning rates were determined as $10^{-4}$ for optimizing the parameter $\theta_{\mathrm{f}}$, $2\times10^{-4}$ for the parameters $ \theta_{\rm enc}$, $\theta_{\rm agg}$ and $\theta_{\rm clf}$, and $4\times10^{-6}$ for the parameter $\theta_{\mathrm{g}}$ in which the learning rate was multiplied by 0.1 every three epochs.
Random horizontal flip and random rotations ($0^\circ, 90^\circ, 180^\circ, 270^\circ$) were applied to the input image patches as the data augmentation.
All the parameters of the model were simultaneously optimized in the above setting.
It took about 20 hours to perform five-fold cross-validation by a computer with eight Quadro RTX 5000 (NVIDIA, U.S.).
Our source code is available from https://github.com/PersAM-MIL/PersAM.


\subsection{Subtype classification}

We performed the three-class classification experiment using the dataset outlined above.
\paragraph{Baseline methods}
The proposed PersAM model was compared with the following baseline models:
\begin{enumerate}
\item MLP using clinical factors as input (clinical MLP)\\
Clinical MLP employs a three-layer NN that has hidden layers with 256 and 512 units and uses a 28-dimensional vector indicating clinical factors as input data.
The training of clinical MLP was performed in 500 epochs, where the learning rate was set to $10^{-3}$ without scheduling.

\item Attention-based MIL using images \cite{ilse2018attention} (img MIL)\\
Img MIL employs an attention-based MIL that aggregates 2048-dimensional feature vectors in a bag and predicts the class label from an aggregated feature vector using the classifier $g_{\textrm{clf}}$ with the hidden layer having 1024 units.

\item Attention-based MIL using images and clinical factors (img-clinical MIL)\\
In img-clinical MIL, in addition to a 512-dimensional aggregated feature vector computed from image patches by the attention-based MIL, a 28-dimensional clinical factor is also used as an input for computing the 512-dimensional feature.
By concatenating the aggregated feature vector for images in a bag and the computed feature vector for clinical factors, the classifier $g_{\textrm{clf}}$ predicts the class using the 1024-dimensional concatenated feature vector through the hidden layer with 512 units.

\item Transformer-based MIL using images (img Transformer)\\
In img Transformer, only one class token was concatenated to feature vectors for $L$ image patches.
The classifier $g_{\textrm{clf}}$ predicts the class using the encoded class token (it is a common technique in the Transformer-based classification model).

\item Transformer-based MIL using images and clinical factors (img-clinical Transformer)\\
Similar to img Transformer, img-clinical Transformer uses only one class token, and it is concatenated to feature vectors for $L$ image patches and $M$ clinical factors.
Img-clinical Transformer predicts the class using the encoded class token as an input for the classifier $g_{\textrm{clf}}$.

\end{enumerate}
The setting for an optimization method and learning rates are the same as above except for clinical MLP.

\paragraph{Results}
The classification results are shown in Table \ref{table:acc}, where each row shows the mean accuracy and standard error in three-class classification by five-fold cross-validation.
The results show that the proposed method achieved the highest accuracy compared to all the baseline methods.
In particular, whereas the baseline methods using image and clinical factors showed low accuracy, our proposed method classified the subtype more accurately by properly aggregating image and clinical features through the multimodal aggregator.
%

\begin{table}[t]
 \caption{Comparison of mean accuracy and standard error in three-class classification by five-fold cross-validation. The proposed method achieved the highest classification accuracy.}
 \label{table:acc}
 \centering
 \begin{tabular}{lc}
  \hline
   Method & Accuracy \\
   \hline \hline
   Clinical MLP & 0.5795 $\pm$ 0.0071\\
   \hline
   Img MIL &  0.8195 $\pm$ 0.0090\\
   Img-clinical MIL & 0.8230 $\pm$ 0.0085\\
   Img Transformer & 0.8219 $\pm$ 0.0140\\
   Img-clinical Transformer & 0.8147 $\pm$ 0.0141\\
   \hline
   Proposed &  \textbf{0.8313} $\pm$ 0.0149\\
   \hline
 \end{tabular}
\end{table}

\subsection{Attention visualization}
\begin{figure}[t!]
\begin{center}
\includegraphics[width=1.0\linewidth]{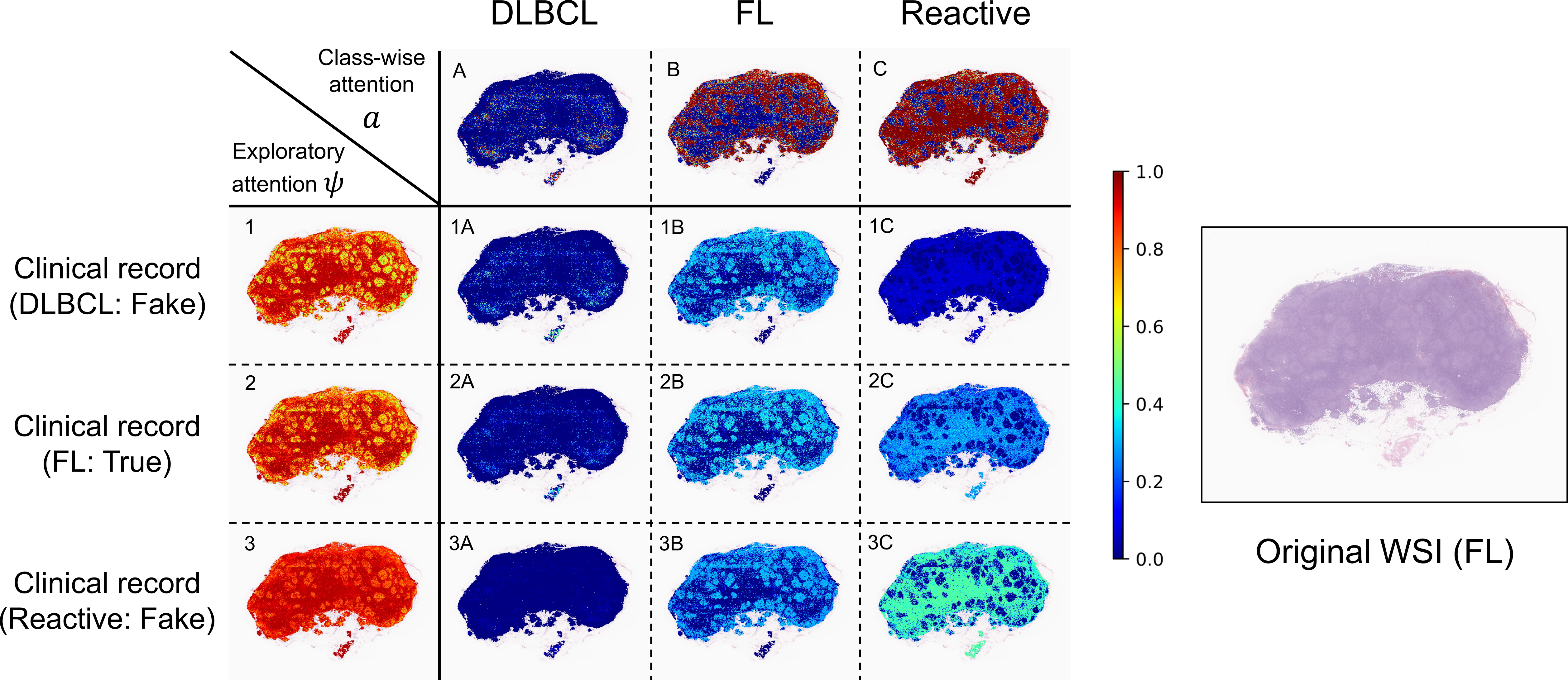}
\end{center}
 \caption{
 Three types of attentions for an FL case.
 The image on the right is a thumbnail of the original WSI.
 Regions with red color in each attention represent attention regions according to the color bar.
 The left column, top row, and each element of the matrix show the class-wise attentions, exploratory attentions, and explanatory attentions, respectively.
 Note that the first (DLBCL) and the third (Reactive) rows are the results when \emph{fake} clinical records are provided for confirming the change of attentions.
 The explanatory attention for FL is enhanced when a clinical record of the FL case is input (2B), and the explanatory attention for Reactive is enhanced when a clinical record of the Reactive case is input (3C), because the case should have some difficulty in the diagnosis using only image.
}
\label{fig:vismat1}
\end{figure}

\begin{figure}[t
 !]
\begin{center}
\includegraphics[width=1.0\linewidth]{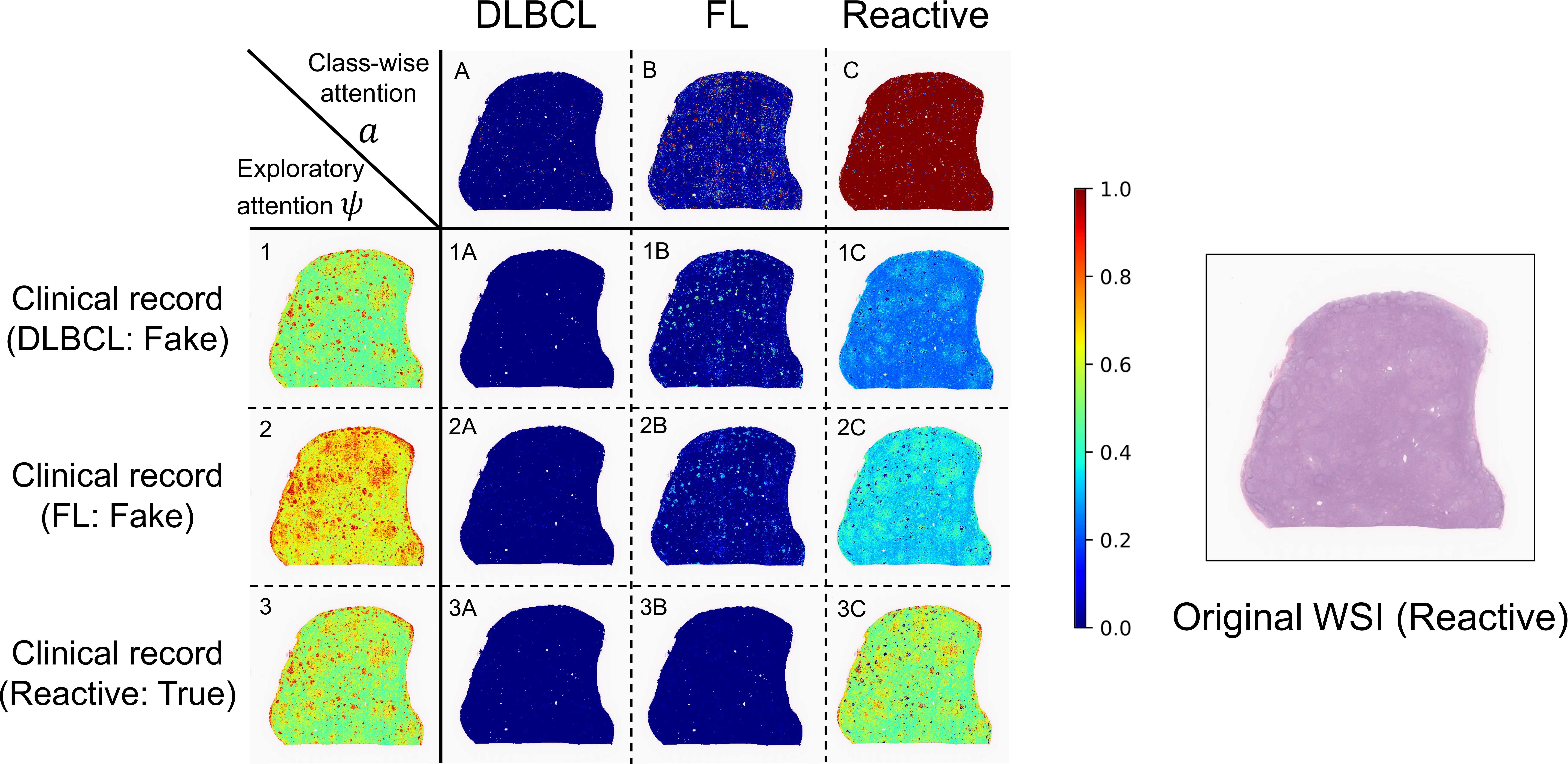}
\end{center}
 \caption{
 Three types of attentions for a Reactive case (see the caption in Fig.~\ref{fig:vismat1}).
 Note that the first (DLBCL) and the second (FL) rows are the results when \emph{fake} clinical records are provided for confirming the change of attentions.
 Class-wise attentions for FL focus on the follicular regions (B) and those for Reactive focus on the outside follicular regions (C). With a Reactive clinical factor, explanatory attention for FL no longer has high values in the follicular region (3B) to strongly predict the subtype as Reactive (3C).
}
\label{fig:vismat2}
\end{figure}

\paragraph{Class-wise, exploratory, and explanatory attentions}
We also performed visualization experiments to demonstrate that the proposed personalized attentions could be adaptively changed according to input clinical records.
In the visualization results, attention weights ranging from zero to one were assigned in the range blue to red.
Figs.~\ref{fig:vismat1} and \ref{fig:vismat2} show the visualization results of class-wise attentions, exploratory attentions, and explanatory attentions, where the images on the right are thumbnails of the original WSIs.
In the matrices on the left, the columns show the class-wise attentions $\{a_{\ell, c}\}_{(\ell, c) \in [L] \times [C]}$, the rows show exploratory attentions $\{\psi_\ell\}_{\ell \in [L]}$, and each element shows explanatory attentions $\{a^{\prime}_{\ell, c}\}_{(\ell, c) \in [L] \times [C]}$, where clinical records sampled from other cases of three different subtypes were input with the WSI of a patient instead of the original clinical factor of that patient.
\emph{Fake} clinical records were used to confirm that exploratory and explanatory attentions changed when different clinical records were input with the same WSI.

Fig.~\ref{fig:vismat1} shows the results for an FL case.
It is known that the follicular structure, a subtype-specific region for FL, is important in the diagnosis of FL cases.
This case has large follicular structures in the tissue, and we can confirm that exploratory attention $\{\psi_{\ell}\}_{\ell \in [L]}$, i.e., the follicular region on which focus is placed, changes depending on the clinical records.
This case should have some difficulty in the diagnosis using only images, and explanatory attentions change according to the input clinical records.

Fig.~\ref{fig:vismat2} shows the result for a Reactive case.
It is known that some Reactive cases have a similar appearance to FL cases.
This case also has small follicular structures in the tissue, which is similar to FL, and exploratory attentions $\{\psi_{\ell}\}_{\ell \in [L]}$ are enhanced in those regions.
Class-wise attentions for FL focus on the follicular regions and those for Reactive focus on the outside follicular regions.
Detailed discussions for these results will be done with magnified image patches later.

The visualization results whose attentions were changed are observed for parts of the dataset,
and not all cases changed their attentions depending on input clinical records.
From the above results, it is expected that the changes of attentions are caused by whether the input WSI is pathologically typical or not;
a case whose disease can clearly be determined only from a WSI does not change its attentions if a different clinical record is input together,
and on the other hand, a case whose WSI has ambiguous features to identify the subtype has the possibility of changes of attentions depending on the input clinical record.
To qualitatively confirm this, an expert hematopathologist
 (one of the authors, who is an institution member with over 15 years of experience diagnosing more than 10,000 cases of lymphoma) investigated whether each case was typical or not in both cases whose attention were changed and not.
%
We targeted FL cases to easily interpret the observation results.
The pathologist observed many WSIs of FL cases that changed attentions and did not change attentions when the different (fake) clinical records were input and evaluated their typicalness of FL with blind whether the attentions of each case changed or not.
A case that was determined as FL only from a WSI was evaluated as a typical FL case,
and a case that cannot be determined as FL only from a WSI and requires immunohistochemical (IHC) stains was evaluated as an atypical case.
The results are discussed in \S4.4.
%
%

\paragraph{Clinical-record-to-patch attentions}

\begin{figure}[t!]
\begin{center}
\includegraphics[width=1.0\linewidth]{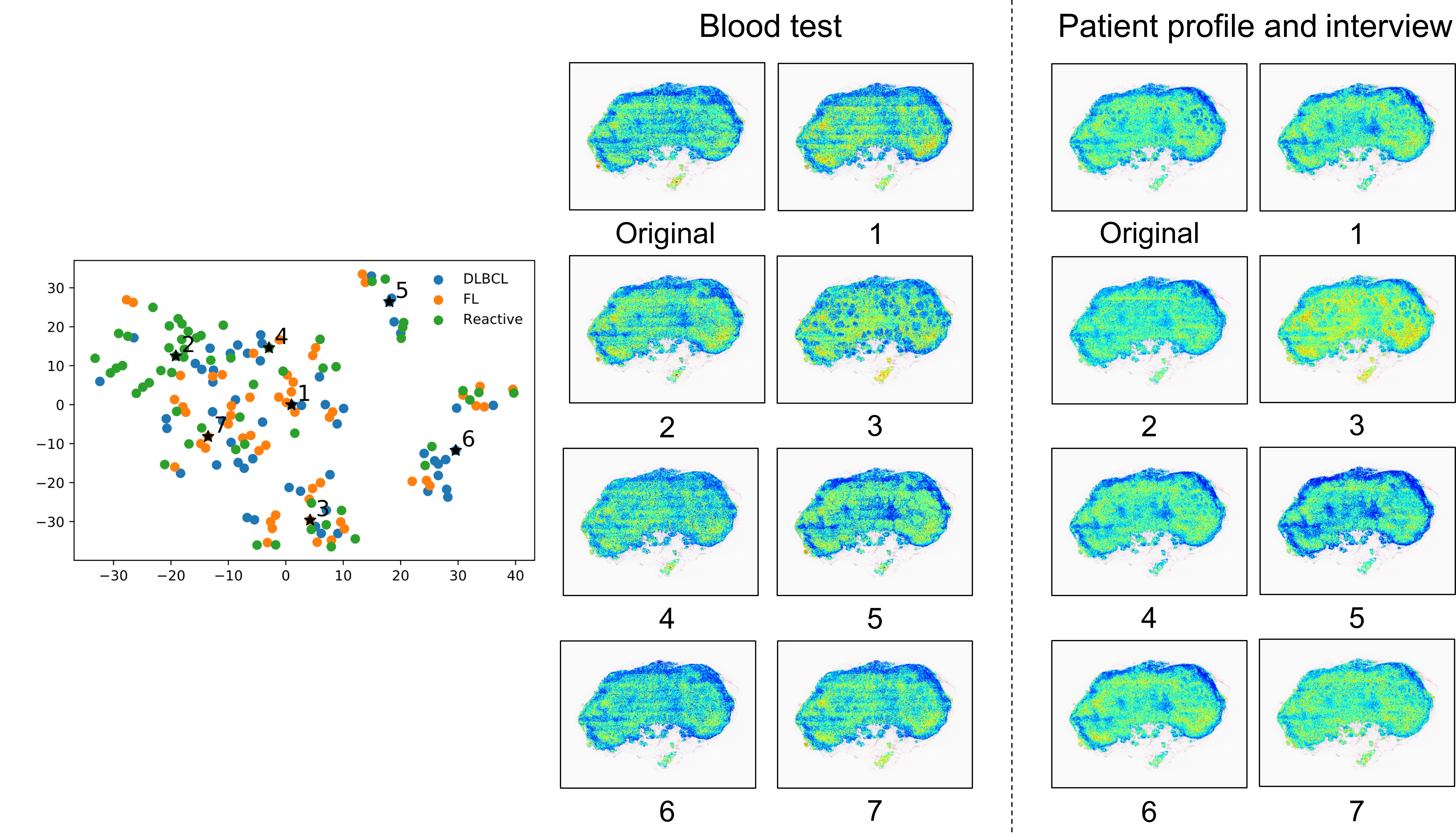}
\end{center}
\caption{
The visualization result of clinical-record-to-patch attentions for an FL case, where the clinical factor of the original case is replaced with those of the representative cases of $k$-medoids clustering result using two-dimensional t-SNE embedded features.
The plots on the left represent the embedded clinical factors, where $\star$ are the representative cases in each cluster.
The images in the middle are the visualization results of clinical-record-to-patch attentions corresponding to blood test for each representative case.
The images on the right are the visualization results of clinical-record-to-patch attentions corresponding to interview for each representative case.
We can confirm that the proposed PersAM could adaptively change clinical-record-to-patch attentions depending on the input clinical records.
}
\label{fig:visself}
\end{figure}

We call self-attentions in the bag representation $\hat{\bm H}$ which indicates attentions from each clinical factor to image patches ``\emph{clinical-record-to-patch attentions}''.
As an additional experiment, we visualized how clinical-record-to-patch attentions changed according to input clinical factors.
Fig. \ref{fig:visself} shows the visualization result of clinical-record-to-patch attentions for an FL case, where the clinical factor of an original (real) case is replaced with the those of representative (fake) case in the clustering results.
\emph{Fake} cases were used to confirm that clinical-record-to-patch attentions changed when different clinical records were input with the same WSI similarly to Figs.~\ref{fig:vismat1} and \ref{fig:vismat2}.

In the clustering, $k$-medoids method was applied to two-dimensional t-SNE embedded features that were calculated from 28-dimensional clinical factors.
We determined the number of clusters by looking at t-SNE embedded features and set to $k=7$.
Instead of the original clinical factors, the clinical factors of the representative cases in each cluster were input with a WSI of the original case into the PersAM model.
The plots on the left represent the embedded clinical factors, in which $\star$ are the representative cases in each cluster.
The images on the middle and right are the visualization results of clinical-record-to-patch attentions corresponding to blood test and interview, respectively.
Attention weights of image patches are normalized in each case for visualization.

We can confirm that the proposed PersAM could adaptively change clinical-record-to-patch attentions depending on the clinical records that were input with an original WSI.
The visualization results for the representative case 1, 2, 4, and 7 are similar to each other because the embedded clinical factors were located close, but the result for case 3 focuses on the follicular structures and the result for case 5 focuses on the outside follicular structures.
We confirmed that clinical-record-to-patch attention also changed depending on clinical factors by effectively encoding the relationship between image patches and clinical factors.

\subsection{Pathological viewpoint on attention}

\begin{figure}[t!]
\begin{center}
\includegraphics[width=1.0\linewidth]{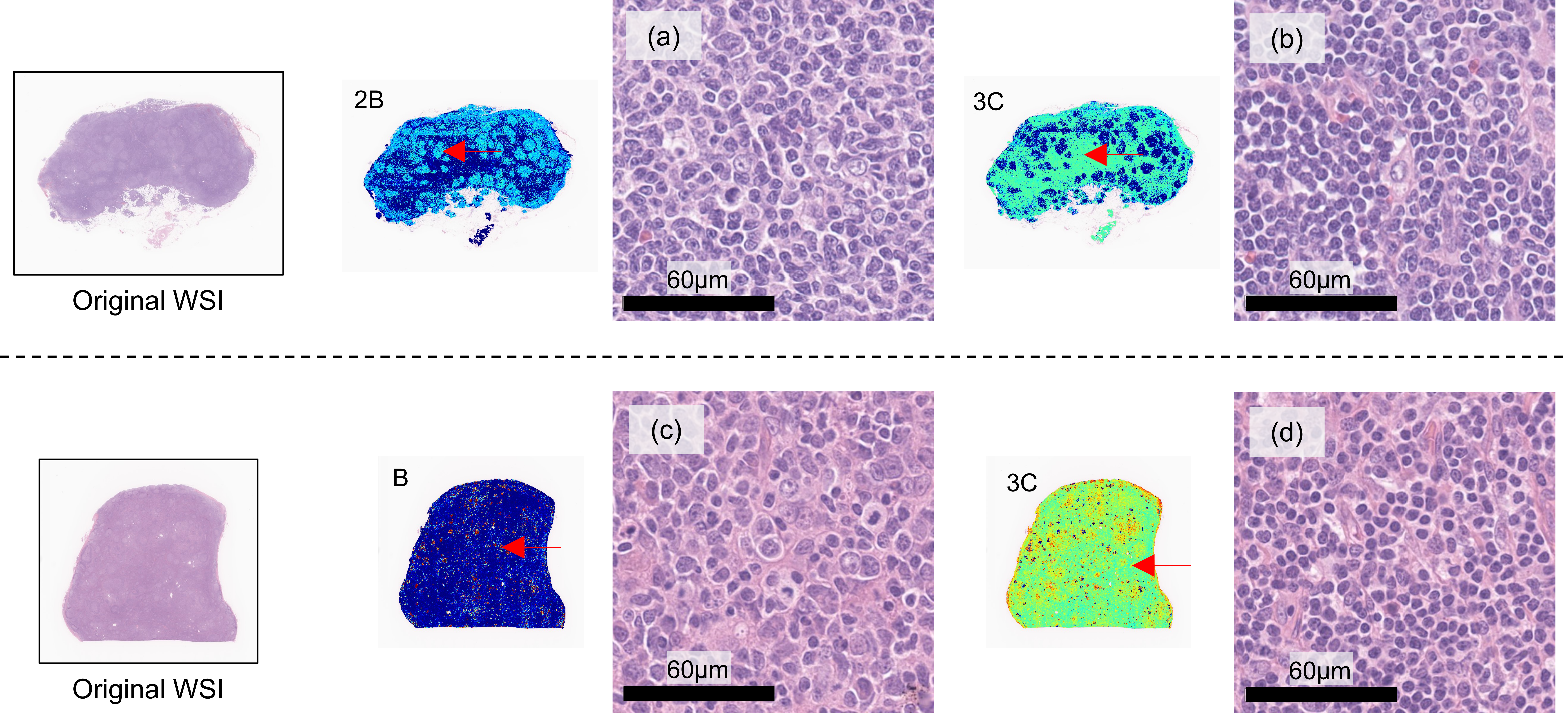}
\end{center}
 \caption{
The magnified image patches of attention regions in Figs.~\ref{fig:vismat1} and \ref{fig:vismat2}.
(a) and (b) are high-resolution image patches in Fig.~\ref{fig:vismat1}(2B) and Fig.~\ref{fig:vismat1}(3C).
(c) and (d) are high-resolution image patches in Fig.~\ref{fig:vismat2}(B) and Fig.~\ref{fig:vismat2}(3C).
Changed attention regions depending on different clinical records show typical image patches as seen in the corresponding subtype.
}
\label{fig:mag}
\end{figure}

Here we discuss the detailed results for the experiment of attention visualization with the expert hematopathologist's comments.
For Fig.~\ref{fig:vismat1}, the hematopathologist made a comment on this result that the change in the explanatory attentions in this case is reasonable because pathologists need to focus more on the follicular regions to identify FL cases (Fig.~\ref{fig:mag}(a)) and on the outside follicular regions to identify Reactive cases (Fig.~\ref{fig:mag}(b)).
In general, as mentioned above, follicular region is important to identify FL cases, but the case of Fig.~\ref{fig:vismat1} has a lot outside follicular regions compared to typical FL cases.%
When a case has a large part of such outside follicular regions in the WSI, it is expected that the classification model focuses on outside follicular regions when the clinical factor of Reactive cases was input with the WSI.

For Fig.~\ref{fig:vismat2}, the hematopathologist made a comment on this result that the change in the explanatory attentions in this case is also reasonable because the model focuses less on the follicular structure to identify Reactive cases.
This case has follicular regions as shown in Fig.~\ref{fig:mag}(c), where histological features of the entire tissue specimen were not typical Reactive case.
In such cases, the pathologist can not identify Reactive case only from a WSI with confidence even if the outside follicular regions has typical features of Reactive cases (Fig.~\ref{fig:mag}(d)).

\begin{figure}[t!]
\begin{center}
\includegraphics[width=0.8\linewidth]{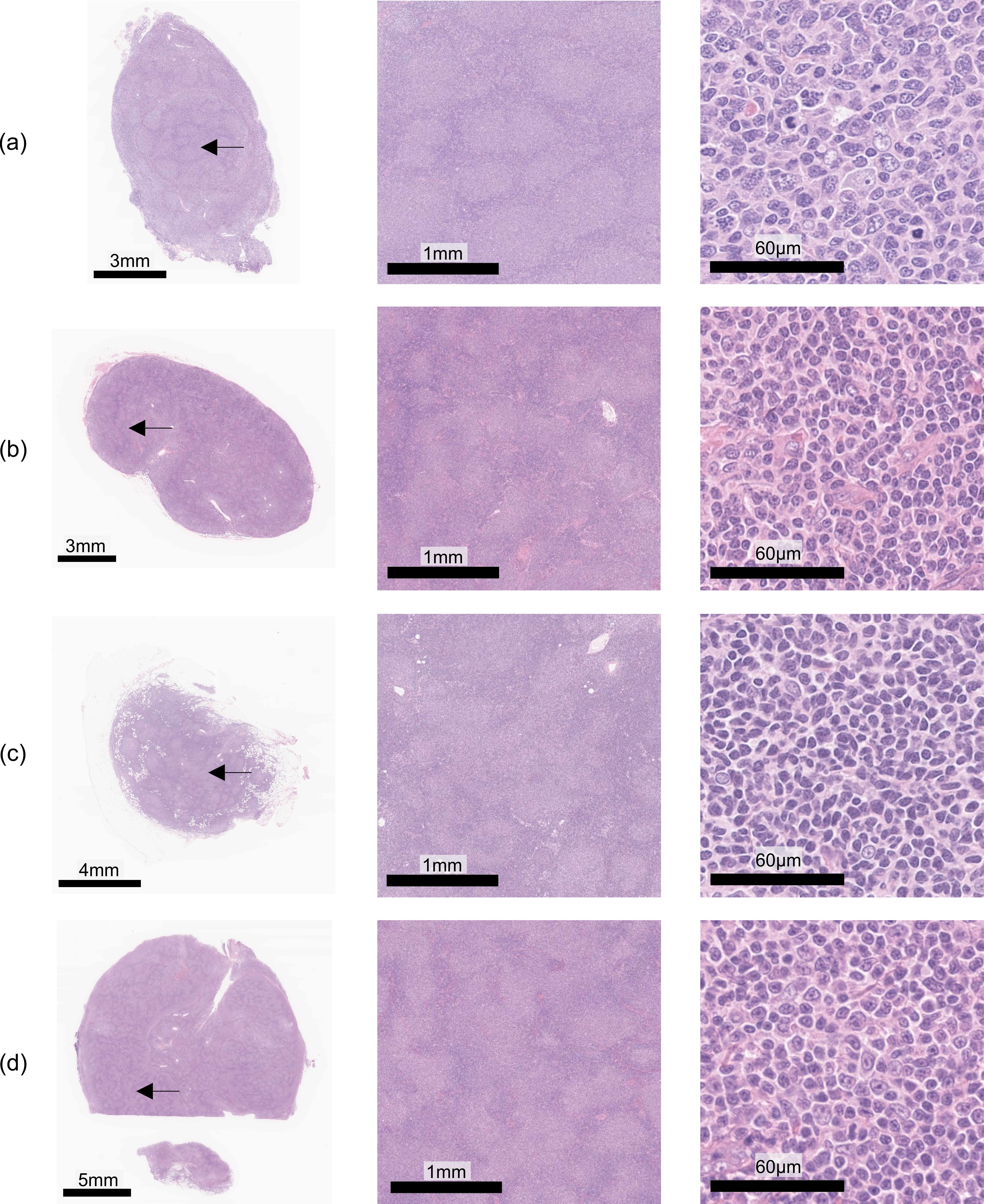}
\end{center}
 \caption{
 Low and high-resolution image patches of the cases evaluated as typical FL cases.
 The attention of all the cases did not change when the different clinical factors were input with the original WSIs in the attention visualization.
 All WSIs have typical FL features and the pathologist can identify them as FL cases only from WSIs.
}
\label{fig:sbj1}
\end{figure}

\begin{figure}[t!]
\begin{center}
\includegraphics[width=0.8\linewidth]{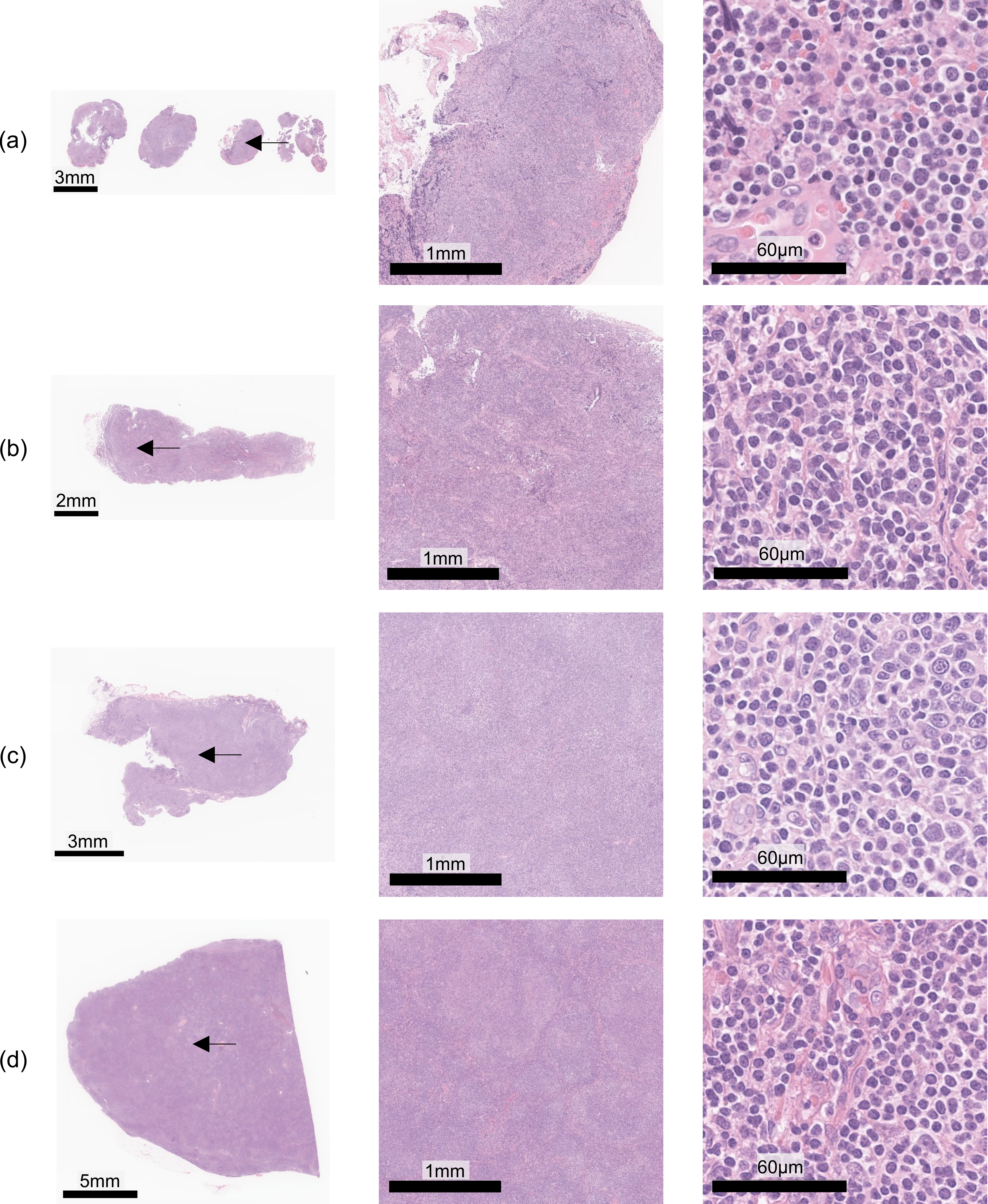}
\end{center}
 \caption{
 Low and high-resolution image patches of the cases evaluated as atypical FL cases.
 The attention of all the cases changed when the different clinical factors were input with the original WSIs in the attention visualization.
 All WSIs do not have definitive features to identify them as FL.
}
\label{fig:sbj2}
\end{figure}

Furthermore, we discuss the results of the investigation of typicalness with magnified images in Figs.~\ref{fig:sbj1} and \ref{fig:sbj2}.
Figure~\ref{fig:sbj1} shows cases, which were evaluated as typical FL cases by the pathologist, in the cases where attentions did not change regardless of input clinical records.
All these cases were evaluated as typical FL cases since follicular regions exist in the entire tissue specimens, which enables pathologists to identify them as FL cases only from WSIs.
Figure~\ref{fig:sbj2} shows cases, which were evaluated as atypical FL cases by the pathologist, in the cases where attentions changed depending on input clinical records.
Most cases have less follicular regions in low magnification and the pathologist can not determine them as FL cases due to the lack of definitive FL features.
In Fig.~\ref{fig:sbj2}(d), there are a lot of nodes in the tissue specimen, and the pathologist expects other diseases and has to require IHC stains.
The proposed PersAM method can provide a reasonable explanation that is similar to pathologists' decision-making where the subtype of typical cases can be identified only from tissue specimens regardless of clinical records and attention regions of atypical cases are affected by clinical records.

\section{Conclusion}
In this study, to develop an AI system that mimics the diagnosis process of human pathologists, we proposed the PersAM method, which adaptively changes the attention regions according to patient clinical records.
Our proposed method provided three types of attention regions, which were calculated considering the relationship among multimodal information.
The results of experiments conducted with 842 malignant lymphoma cases verify the effectiveness of the PersAM method.

\clearpage

{\small
\bibliographystyle{ieee}
\bibliography{ref}
}

\end{document}